\documentclass[12pt]{iopart}

\usepackage{setstack}
\usepackage{iopams}
\eqnobysec

\usepackage{graphicx}

\makeatletter
\def\@fnsymbol#1{^{\thefootnote}\relax}
\makeatother
\newcommand{\be}{\begin{equation}}
\newcommand{\ee}{\end{equation}}

\begin{document}

\title{The 3-edge-colouring problem on the 4-8 and 3-12 lattices}

\author{J. O. Fj{\ae}restad}
\address{The University of Queensland, School of Mathematics and Physics, Brisbane, QLD 4072, Australia}
\ead{jof@physics.uq.edu.au}

\date{\today}

\begin{abstract}
We consider the problem of counting the number of 3-colourings of the edges (bonds) of the 4-8 lattice and the 3-12 lattice. These lattices are Archimedean with coordination number 3, and can be
regarded as decorated versions of the square and honeycomb lattice, respectively. We solve these edge-colouring problems in the infinite-lattice limit by mapping them to other models whose solution is known. The colouring problem on the 4-8 lattice is mapped to a completely packed loop model with loop fugacity $n=3$ on the square lattice, which in turn can be mapped to a six-vertex model. The colouring problem on the 3-12 lattice is mapped to the same problem on the honeycomb lattice. The 3-edge-colouring problems on the 4-8 and 3-12 lattices are equivalent to the 3-vertex-colouring problems (and thus to the zero-temperature 3-state antiferromagnetic Potts model) on the 
``square kagome" (``squagome") and ``triangular kagome" lattices, respectively.
\end{abstract}

\vspace{2pc}
\submitto{J. Stat. Mech.}

\maketitle

\section{Introduction}
\label{intro}

Counting problems involving strong local constraints arise in many different contexts. One important class of such problems deals with graph colourings, which is a topic of interest in several  fields, including combinatorics and graph theory \cite{biggs}, theoretical computer science, and statistical mechanics. In the latter field, connections have been identified
between various colouring problems and other types of problems, such as Potts models \cite{wu-potts}, loop models \cite{baxter-70,kh-95}, and tiling 
\cite{tiles} and folding \cite{foldings} problems. Recently quantum generalizations of some colouring problems have also been studied~\cite{castelnovo}.

If each vertex of a graph $G$ is coloured with one out of $q$ colours, such that no vertices connected by an edge have the same colour, the result is said to be
a $q$-vertex-colouring of $G$.\footnote{Vertices and edges are often instead called sites and bonds in the physics literature.} The number $P_G(q)$ of such colourings for a graph with $N$ vertices is a polynomial in $q$ of degree $N$, called the chromatic polynomial of $G$, and the smallest integer $q$ for which $P_G(q)>0$ is the chromatic number $\chi_G$ \cite{biggs}. The problem of determining $P_G(q)$ for a given $q$ is the $q$-vertex-colouring problem on $G$. For $q\geq \chi_G$, $P_G(q)$ is equal to the zero-temperature partition function of the $q$-state antiferromagnetic (AF) Potts model on $G$. These AF Potts models are, for sufficiently large $q$, examples of \textit{un-}frustrated systems with an extensive ground state entropy: the ground state degeneracy increases exponentially with $N$, while the energy associated with each edge takes the lowest value possible.

Similar considerations apply to colourings of the edges of a graph. If each edge of a graph $G$ is coloured with one out of $q$ colours, such that no edges meeting at a vertex
have the same colour, the result is said to be a $q$-edge-colouring of $G$. The number of such colourings is the edge chromatic polynomial $P_G^{(e)}(q)$, and the edge chromatic number $\chi_G^{(e)}$ is the smallest integer $q$ such that $P_G^{(e)}(q)>0$. The $q$-edge-colouring problem on $G$ is equivalent to the $q$-vertex-colouring problem on a
related graph $G'$, defined such that every edge of $G$ hosts a vertex of $G'$, and two vertices in $G'$ are connected by an edge if their host edges in $G$ meet at a vertex. Nevertheless, in spite of this equivalence, it may still be preferable to study this problem in its original edge-colouring formulation.

An interesting class of lattices in two dimensions are the Archimedean lattices, which consist of uniform tilings of the plane with regular polygons such that all vertices have an identical environment. There are 11 such lattices, the most familiar ones being the square, honeycomb, triangular, and kagome lattices. The edge chromatic number for an Archimedean lattice
is equal to its coordination number \cite{ecn}. The simplest nontrivial $q$-edge-colouring problem for an Archimedean lattice is therefore obtained by taking $q$ equal to its coordination number. This is the problem we consider in the following. 
Let ${\cal L}$ be a lattice with $N$ vertices and coordination number $q$, with periodic boundary conditions in both directions, so the system is topologically a torus. The number of $q$-edge-colourings $Z_{\cal L}$ scales as $W_{\cal L}^N$ as $N\to\infty$. That is, $W_{\cal L}$ is defined as
\be
W_{\cal L} \equiv \lim_{N\to\infty} Z_{\cal L}^{1/N}.
\label{def-W}
\ee
Previously, $W_{\cal L}$ has been calculated exactly for the 3-edge-colouring problem on the honeycomb lattice \cite{baxter-70} and for the 4-edge-colouring problem on the square lattice \cite{dc-nien-04}. 
In these works the problem was formulated in terms of a transfer matrix that was diagonalized with a coordinate Bethe Ansatz \cite{elab}. 

\begin{figure}[h]
\includegraphics[scale=0.7]{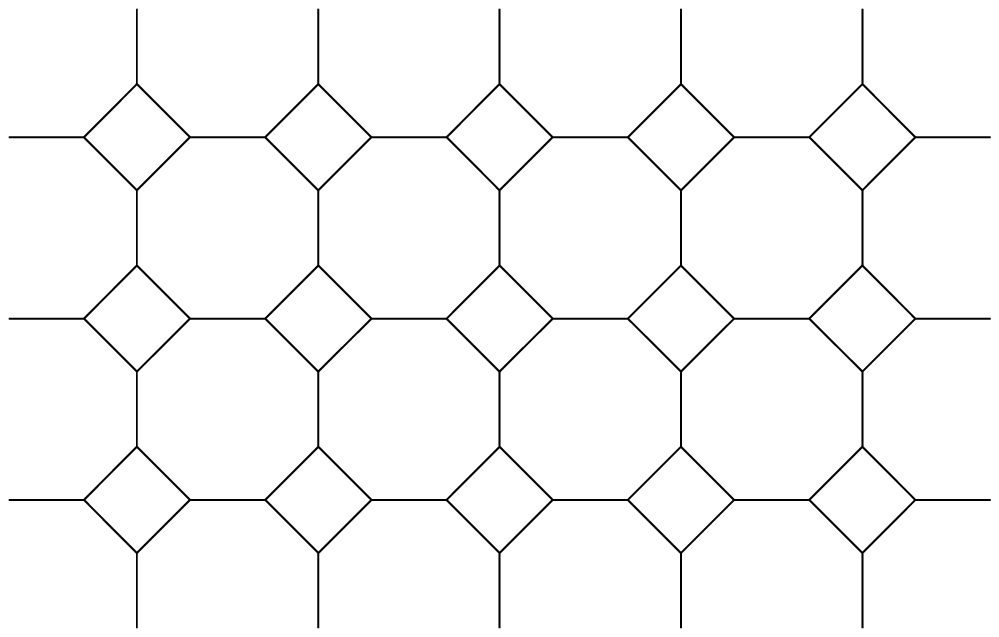}\hspace{1.5cm}
\includegraphics[scale=0.62]{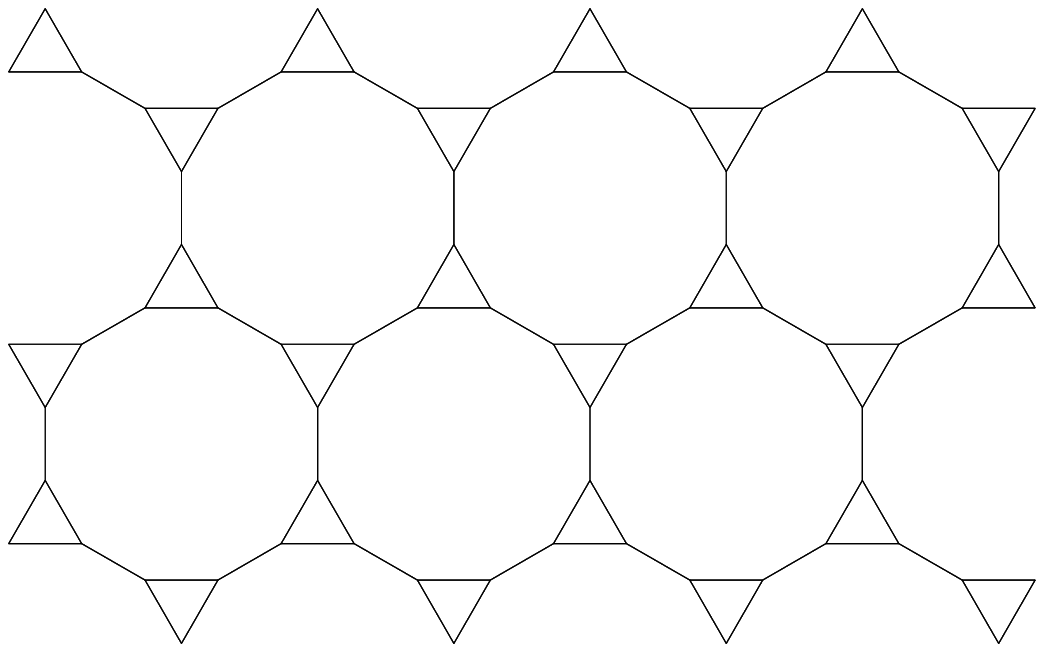}\vspace{0.5cm}\\
\mbox{}\hspace{3.3cm}(a)\hspace{8.0cm}(b)
\caption{(a) The 4-8 lattice. (b) The 3-12 lattice.}
\label{fig:lattices}
\end{figure}

In this paper we calculate $W_{\cal L}$ exactly for the 3-edge-colouring problem on the 4-8 and 3-12 lattices  \cite{alt-name} (see Fig.~\ref{fig:lattices}), both of which are Archimedean lattices with coordination number 3. Another common feature of these lattices is that they can both be regarded as a ``decorated" version of a ``parent" lattice; for the 4-8 (3-12) lattice the parent lattice is the square (honeycomb) lattice. We show that these edge-colouring problems can be mapped to previously solved models on the ``parent" lattices. For the 4-8 lattice the mapping is to a completely packed loop model on the square lattice, which in turn can be mapped to a six-vertex model. For the 3-12 lattice the mapping (which is very simple in this case) is to the 3-edge-colouring problem on the honeycomb lattice. We also briefly discuss the alternative formulation of the 3-edge-colouring problems on the 4-8 and 3-12 lattices as 3-vertex-colouring problems (and thus zero-temperature $3$-state AF Potts models) on the ``square kagome" (``squagome") and "triangular kagome" lattices \cite{kagome-like}, and we comment on the 3-edge-colouring problem on the 4-6-12 lattice, another Archimedean lattice with coordination number 3. 

\section{The 3-edge-colouring problem on the 4-8 lattice}
\label{4-8}

\subsection{Identifying the possible local 3-colouring configurations}
\label{local}

\begin{figure}[h]
\includegraphics[angle=270,scale=0.24]{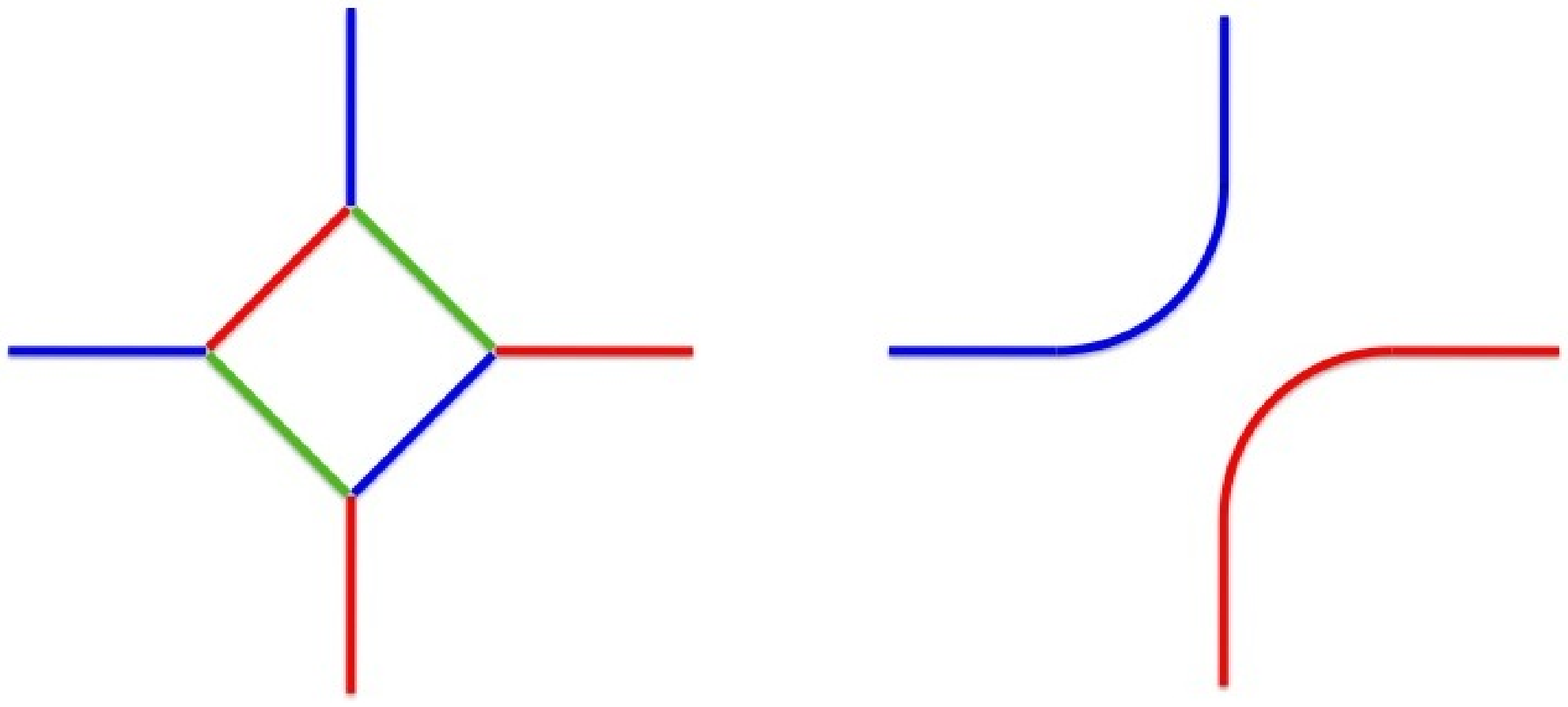}\hspace{0.32cm}
\includegraphics[angle=270,scale=0.24]{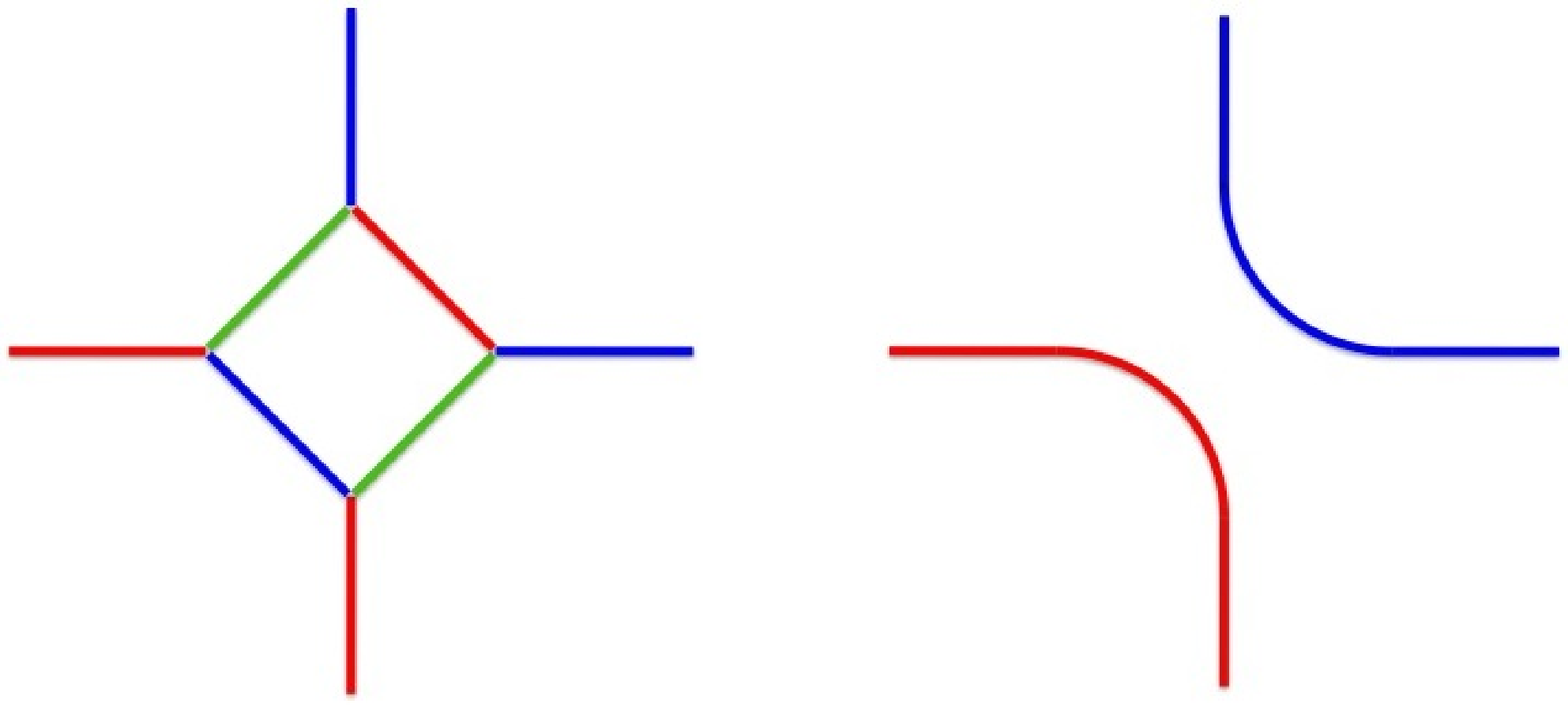}\hspace{0.32cm}
\includegraphics[angle=270,scale=0.24]{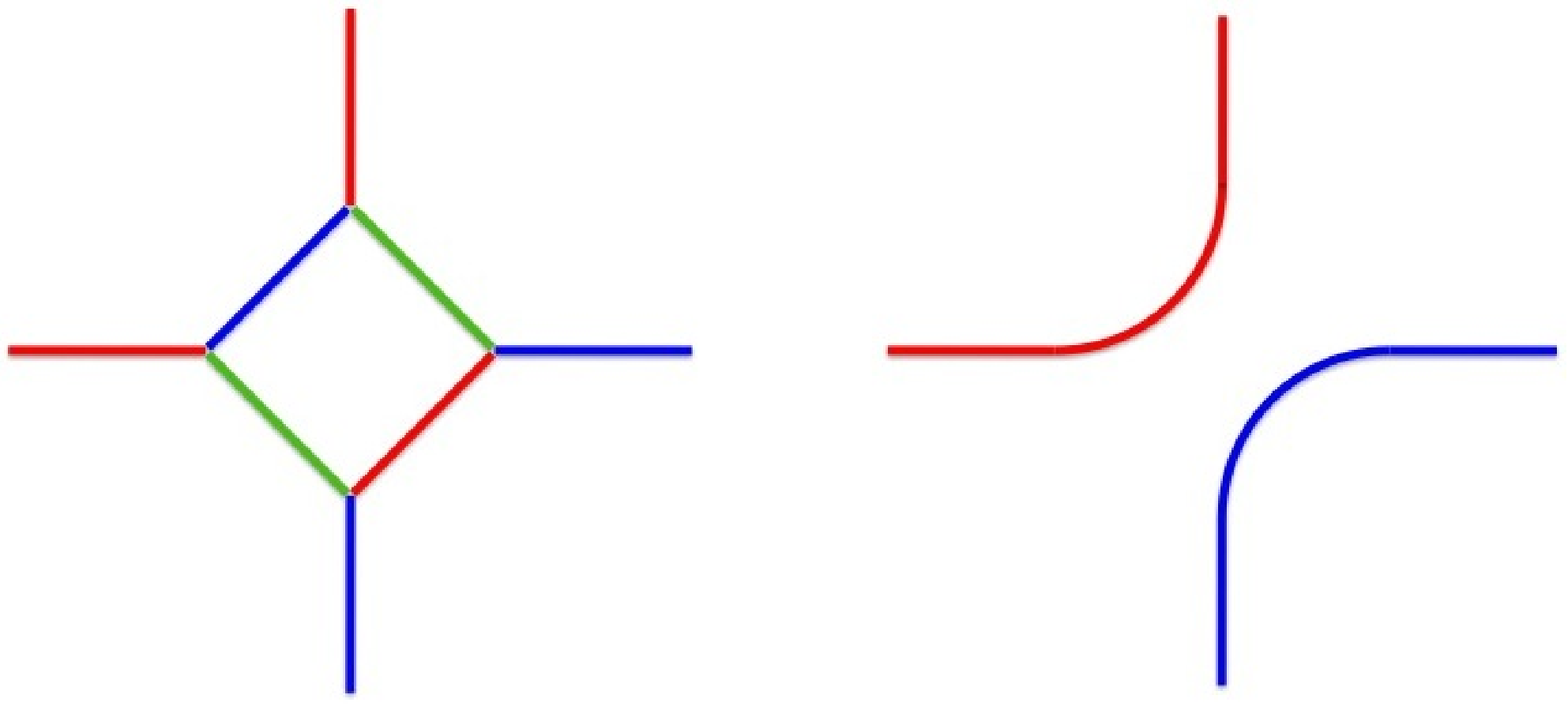}\hspace{0.32cm}
\includegraphics[angle=270,scale=0.24]{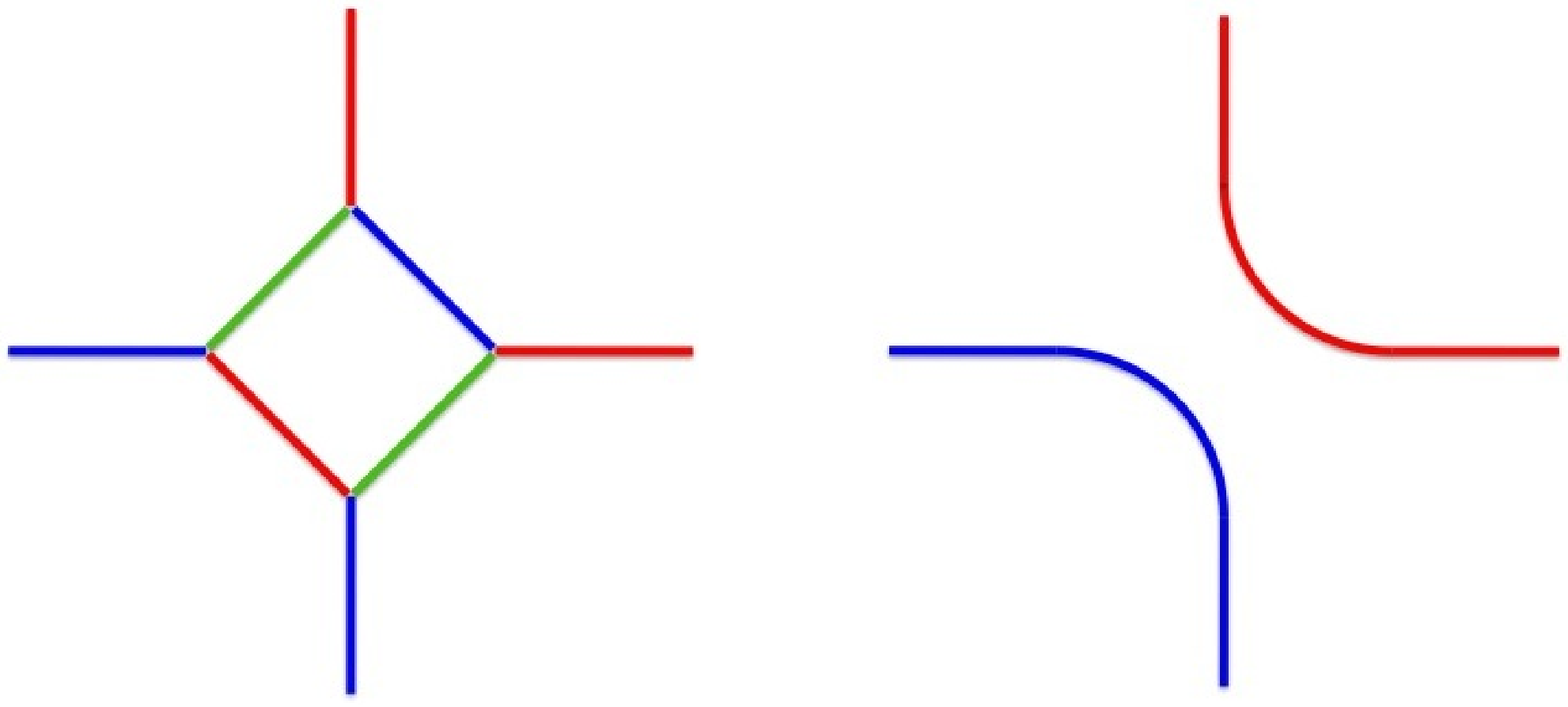}\hspace{0.32cm}
\includegraphics[angle=270,scale=0.24]{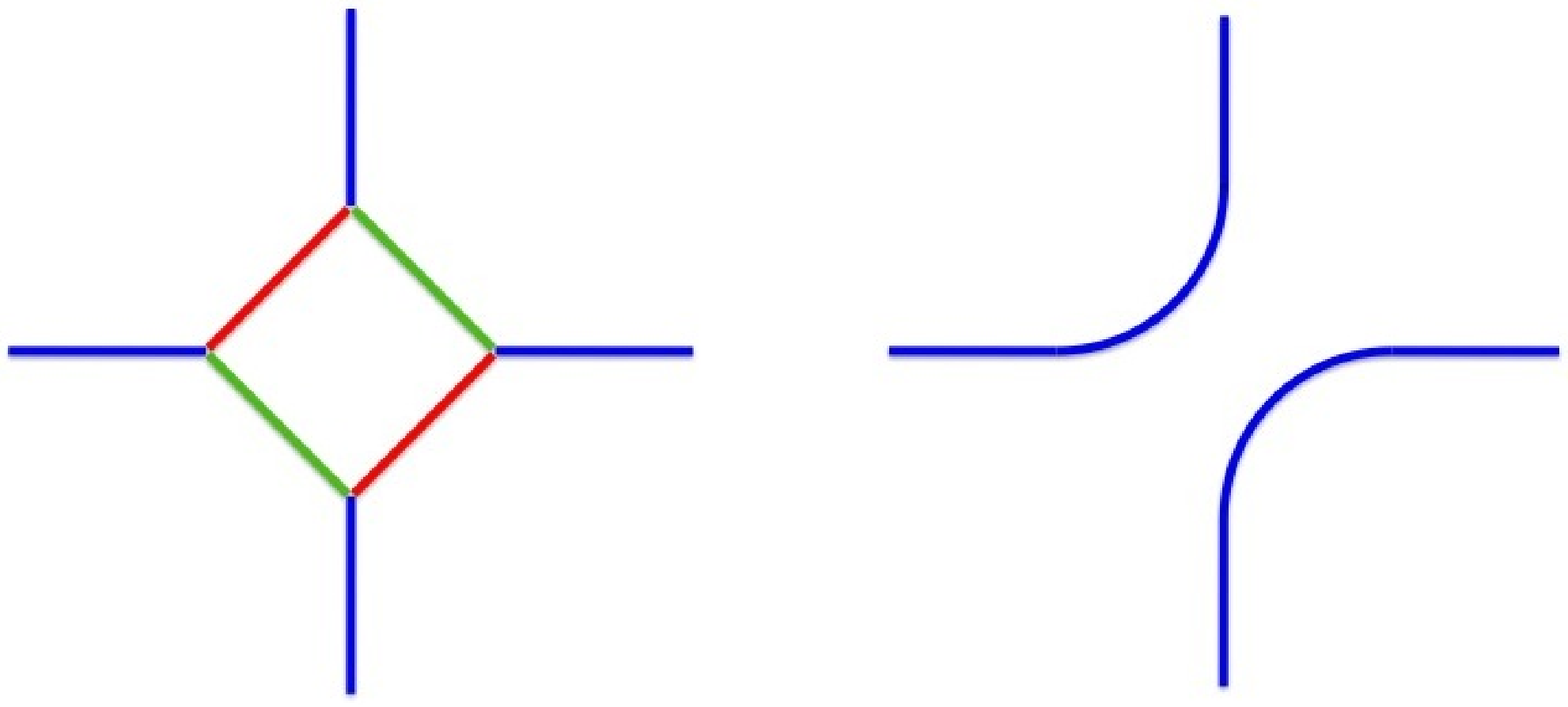}\hspace{0.32cm}
\includegraphics[angle=270,scale=0.24]{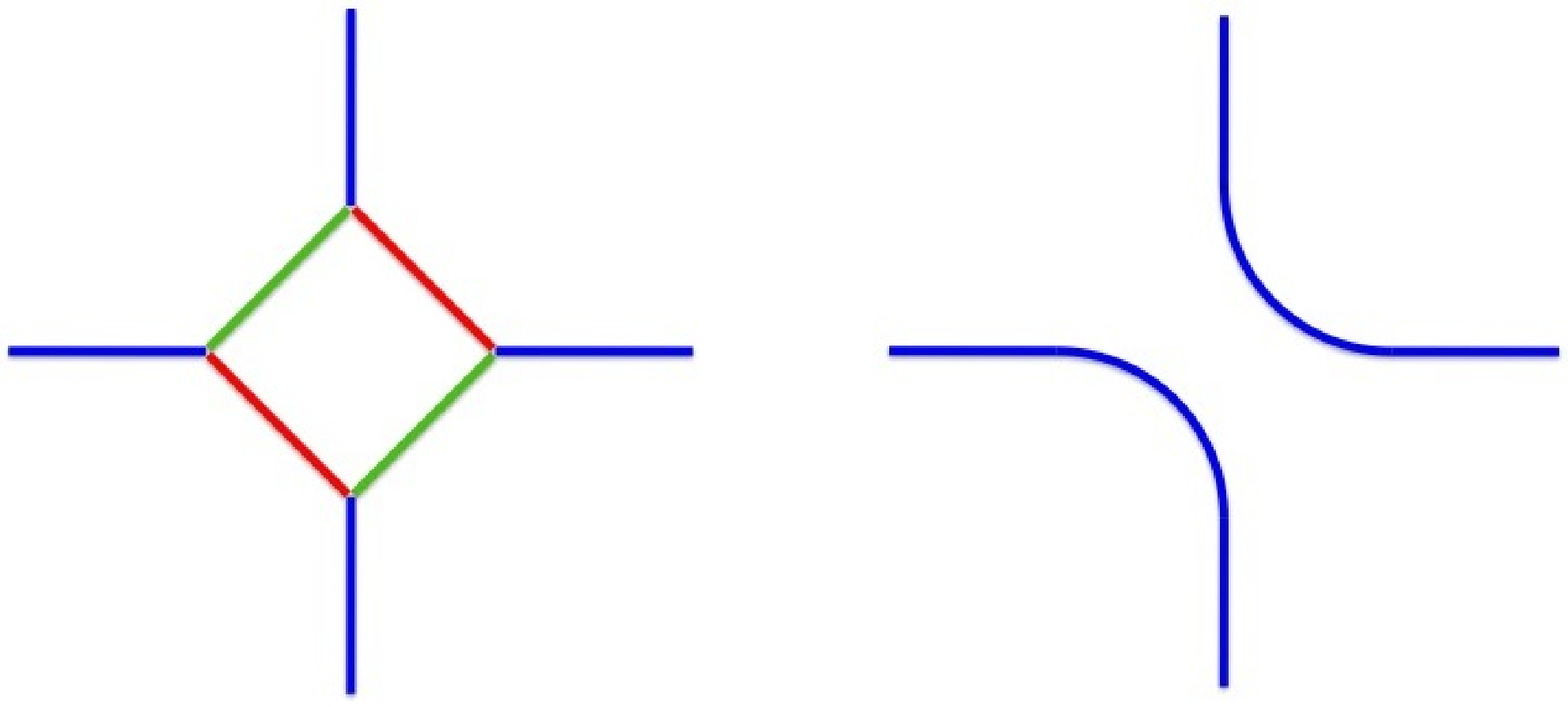}\vspace{0.32cm}\\
\mbox{ }\hspace{0.7cm}(a)\hspace{1.97cm} (b)\hspace{1.97cm} (c)\hspace{1.97cm} (d)\hspace{1.97cm} (e)\hspace{1.99cm} (f) 
\caption{(in colour) Top row: 6 of the 18 possible local 3-colouring configurations on the 4-8 lattice. Bottom row: the corresponding local coloured CPL configurations on the square lattice. The remaining 12 configurations can be obtained by cyclically permuting the 3 colours in the configurations shown here. We have here ordered the colours cyclically as blue $\rightarrow$ red $\rightarrow$ green $\rightarrow$ blue (see text for further details).}
\label{colourings}
\end{figure}

The 4-8 lattice is shown in Fig. \ref{fig:lattices}(a). It can be regarded as a ``decorated" square lattice, the decoration procedure consisting of replacing every vertex on
the square lattice with a square rotated by 45 degrees. We will denote the edges of squares on the 4-8 lattice as ``inner" edges and the edges connecting neighboring squares as ``outer" edges. Consider now one of these squares together with the four outer edges touching it, i.e. a collection of eight edges in total. By considering all possible ways of 3-colouring the four outer edges, and then for each such way considering 3-colourings of the four inner edges, we find that there are only 18 possible (i.e. 
allowed) 3-colourings of the eight edges. In 12 of these colourings, each of the horizontal outer edges is ``paired" with a vertical outer edge in the sense that edges in the same pair have the same colour and edges in different pairs have different colour. For each such colouring of the outer edges there is
only one way to colour the inner edges. Examples of these colourings are shown in Fig.~\ref{colourings}, top row, (a)-(d). In the remaining 6 colourings, all four outer edges have the same colour. For each colour of the outer edges there are then two possible ways to colour the inner edges. Examples of these colourings are shown in Fig. \ref{colourings}, top row, (e)-(f). (Note that the 12 colourings not shown in Fig. \ref{colourings} can be generated by cyclically permuting the 3 colours in the 6 colourings that are shown.)

\subsection{Mapping to a completely packed loop (CPL) model on the square lattice}
\label{cpl}

\begin{figure}[h]
\hspace{5cm}
\includegraphics[scale=0.3]{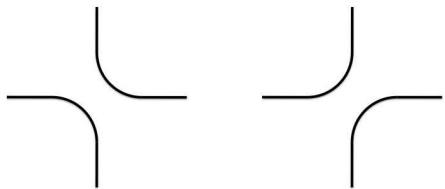}
\caption{The two possible CPL configurations at a vertex on the square lattice.}
\label{CPL-vertices}
\end{figure}

The partition function of the completely packed loop (CPL) model with loop fugacity $n$ on the square lattice is given by
\be
{\cal Z}(n)=\sum_c n^{{\cal N}_c}.
\label{CPL-pf}
\ee
Here the sum is over all configurations of completely packed loops, and ${\cal N}_c$ is the number of loops in configuration $c$. By definition, a CPL configuration consists of non-intersecting loops such that every edge on the square lattice is covered by one loop segment \cite{nien-lh08}. This leads to two possible arrangements of loop segments meeting at a vertex, as shown in Fig. \ref{CPL-vertices}. 

The 3-edge-colouring problem on the 4-8 lattice can be mapped to a CPL model with loop fugacity $n=3$ on the square lattice. The mapping is established by showing that each of the 18 possible local 3-colouring configurations can be mapped to a local CPL configuration of the types shown in Fig. \ref{CPL-vertices}, but with loops that can take 3 possible colours (note that the number of such coloured local CPL configurations is indeed $18=2\cdot 3^2$). Let us first consider the 12 colourings with two different colours on the outer edges. By forgetting about the inner edges (as noted above, their colouring
is uniquely determined in this case and thus represents redundant information) and simply connecting outer edges with the same colour, we get CPL arrangements in which the two meeting loop segments have different colours. This is illustrated in Fig. \ref{colourings} (a)-(d). On the other hand, for the 6 colourings in which the outer edges have the same colour, there are two possible ways to colour the inner edges, which again can be mapped to the two ways that loop segments in the CPL model (now of the same colour, namely that of the outer edges) meet at a vertex. For example, one can make the rule that in this case two outer edges are to be connected if the inner edge between them in the 3-colouring has a colour that follows cyclically after the colour of the outer edges. This is illustrated in Fig. \ref{colourings} (e)-(f). Thus we arrive at a model of completely packed loops in which each loop can take 3 possible colours. It follows that $Z_{\rm{4-8}}$, the number of 3-edge-colourings on the 4-8 lattice, equals the number of loop configurations in this coloured CPL model. This latter number is in turn equal to the partition function of a CPL model of un-coloured loops in which each loop is given a weight of 3, i.e.,
\be
Z_{\rm{4-8}}={\cal Z}(3).
\label{colourings-loops}
\ee
We note that the CPL model (\ref{CPL-pf}) is in a non-critical phase for $n=3$ (the model is critical for $|n|<2$) \cite{nien-lh08}. 

\subsection{Mapping the CPL model to a six-vertex model. Calculation of $W_{\rm{4-8}}$}

The CPL model can be mapped to a six-vertex model \cite{bkw-76,zinn-justin}. To this end, we first rewrite ${\cal Z}(n)$ as a sum over configurations of \textit{directed} loops. 
Writing $n\equiv \omega+\omega^{-1}$, a directed loop is assigned a weight $\omega^{\pm 1}$ where the two signs correspond to the two possible directions of the loop. Next one
writes these directed loop weights as products of local weights associated with each vertex visited by the loop: the assigned vertex weight is $\omega^{1/4}$ or $\omega^{-1/4}$ depending on whether the directed loop turns right or left at the vertex. There is however a subtlety here \cite{bkw-76,zinn-justin}. While the mapping to local vertex weights gives the correct loop weight $\omega+\omega^{-1}=n$ for all contractible loops, on a torus one can also have non-contractible loops (i.e. loops that wind around the torus), and these get an incorrect weight of $1+1=2$ from this procedure. This corresponds to a modified loop model with partition function
\be
\tilde{\cal Z}(n) \equiv \sum_c n^{{\cal N}_c^{\rm{c}}}2^{{\cal N}_c^{\rm{nc}}},
\label{modified}
\ee
where ${\cal N}_c^{\rm{c}}$ (${\cal N}_c^{\rm{nc}}$) is the number of contractible (non-contractible) loops in configuration $c$. We then have
\be
{\cal Z}(n) = \tilde{\cal Z}(n) \cdot \left\langle \left(\frac{n}{2}\right)^{{\cal N}^{\rm{nc}}} \right\rangle,
\label{Z-from-Z0}
\ee
where the expectation value is taken with respect to the model defined by (\ref{modified}). Let the square lattice have $M_x$ ($M_y$) vertices in the $x$ ($y$) direction, i.e. $M=M_x M_y$ vertices in total. Without loss of generality we take $M_y\leq M_x$. The maximum number of non-contractible loops a loop configuration can possibly contain is then $M_x$. Such a configuration would have no contractible loops, while the $M_x$ non-contractible loops, each of length $2M_y$, would wind around the torus in the $y$ direction. This implies that $(n/2)^{M_x}$ is an upper bound for the expectation value in (\ref{Z-from-Z0}). Furthermore, 1 is clearly a lower bound. These bounds imply that for the 4-8 lattice with $N=4M$ sites, $W_{\rm{4-8}}\equiv \lim_{N\to\infty}Z_{\rm{4-8}}^{1/N}$ satisfies 
\be
\hspace{-1.5cm}\lim_{M\to\infty}\tilde{\cal Z}(3)^{1/(4M)} \leq W_{\rm{4-8}} \leq \lim_{M\to\infty}\left[\tilde{\cal Z}(3)\cdot \left(\frac{3}{2}\right)^{M_x}\right]^{1/(4M)} = \lim_{M\to\infty}\tilde{\cal Z}(3)^{1/(4M)}
\ee
(here $\lim_{M\to\infty}\equiv \lim_{M_x,M_y\to\infty}$), from which it follows that
\be
W_{\rm{4-8}}=\lim_{M\to\infty}\tilde{\cal Z}(3)^{1/(4M)}.
\ee

$\tilde{\cal Z}(n)$ can be interpreted \cite{bkw-76,zinn-justin} as the partition function of a six-vertex model on the square lattice with periodic boundary conditions in both directions and model parameters given by
\be
\hspace{-1.5cm} a=b=1,\quad c = \omega^{1/2}+\omega^{-1/2} = \sqrt{n+2} \quad \Rightarrow \quad \Delta \equiv \frac{a^2+b^2-c^2}{2ab}=-\frac{n}{2}.
\ee
Thus $\Delta=-3/2$ for our colouring problem, so the six-vertex model is in the non-critical ``antiferroelectric" phase that occurs for $\Delta < -1$. In this phase the free energy per vertex (site), $f$, is given by \cite{baxter-book}
\be
-\frac{f}{k_B T} = \frac{\lambda}{2} + \sum_{m=1}^{\infty}\frac{e^{-m\lambda}}{m}\frac{\sinh(m\lambda)}{\cosh(m\lambda)},
\label{f-over-kBT}
\ee
where the parameter $\lambda>0$ is defined by $\Delta = -\cosh \lambda$. Introducing $x \equiv e^{-\lambda}$ we find that
\be
e^{-f/k_B T} = \frac{x^{-1/2}}{1-x}\prod_{k=1}^{\infty}\left(\frac{1-x^{4k-1}}{1-x^{4k+1}}\right)^2.
\ee
For $\Delta=-3/2$, $x=2-\varphi$ where $\varphi=(1+\sqrt{5})/2$ is the golden ratio, which implies $x^{1/2}(1-x)=x$. Thus $W_{\rm{4-8}}=(e^{-f/k_B T})^{1/4}$ is given by
\be
W_{\rm{4-8}} = x^{-1/4} \prod_{k=1}^{\infty}\left(\frac{1-x^{4k-1}}{1-x^{4k+1}}\right)^{1/2} = 1.24048\ldots
\label{W-res}
\ee

\section{The 3-edge-colouring problem on the 3-12 lattice}
\label{3-12}

\begin{figure}[h]
\hspace{1cm}
\includegraphics[scale=0.35]{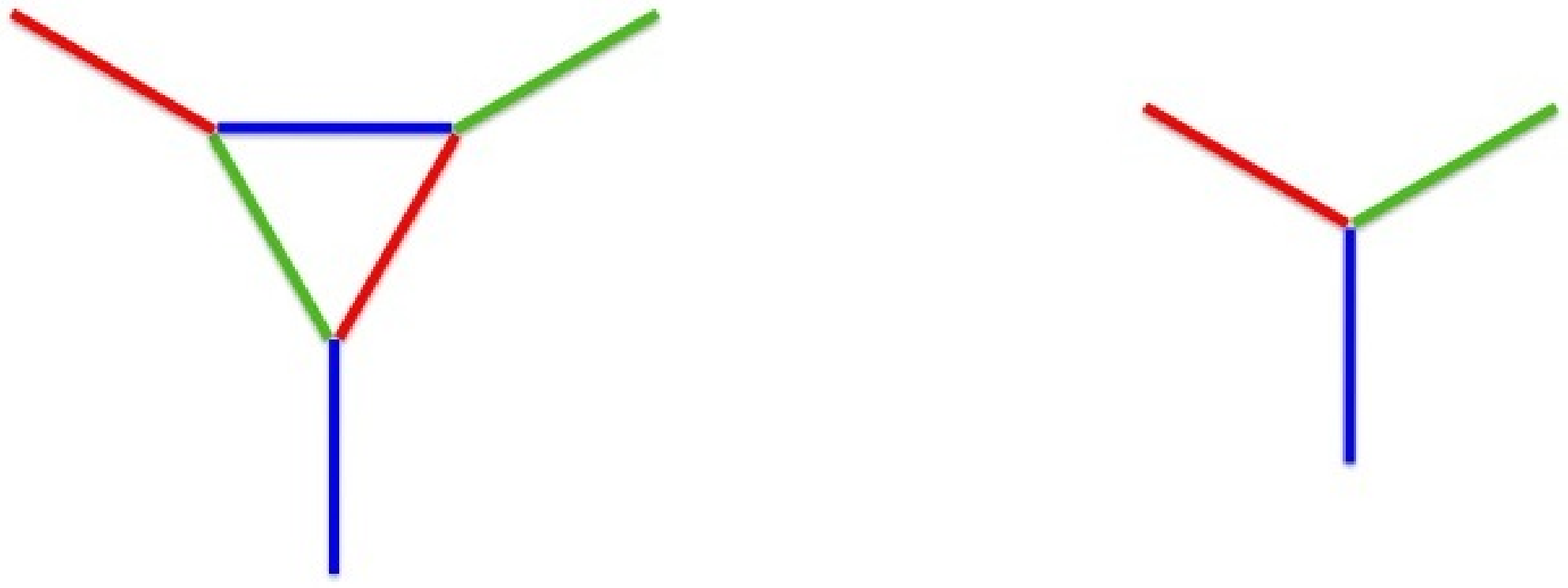}\hspace{2.5cm}
\includegraphics[scale=0.15]{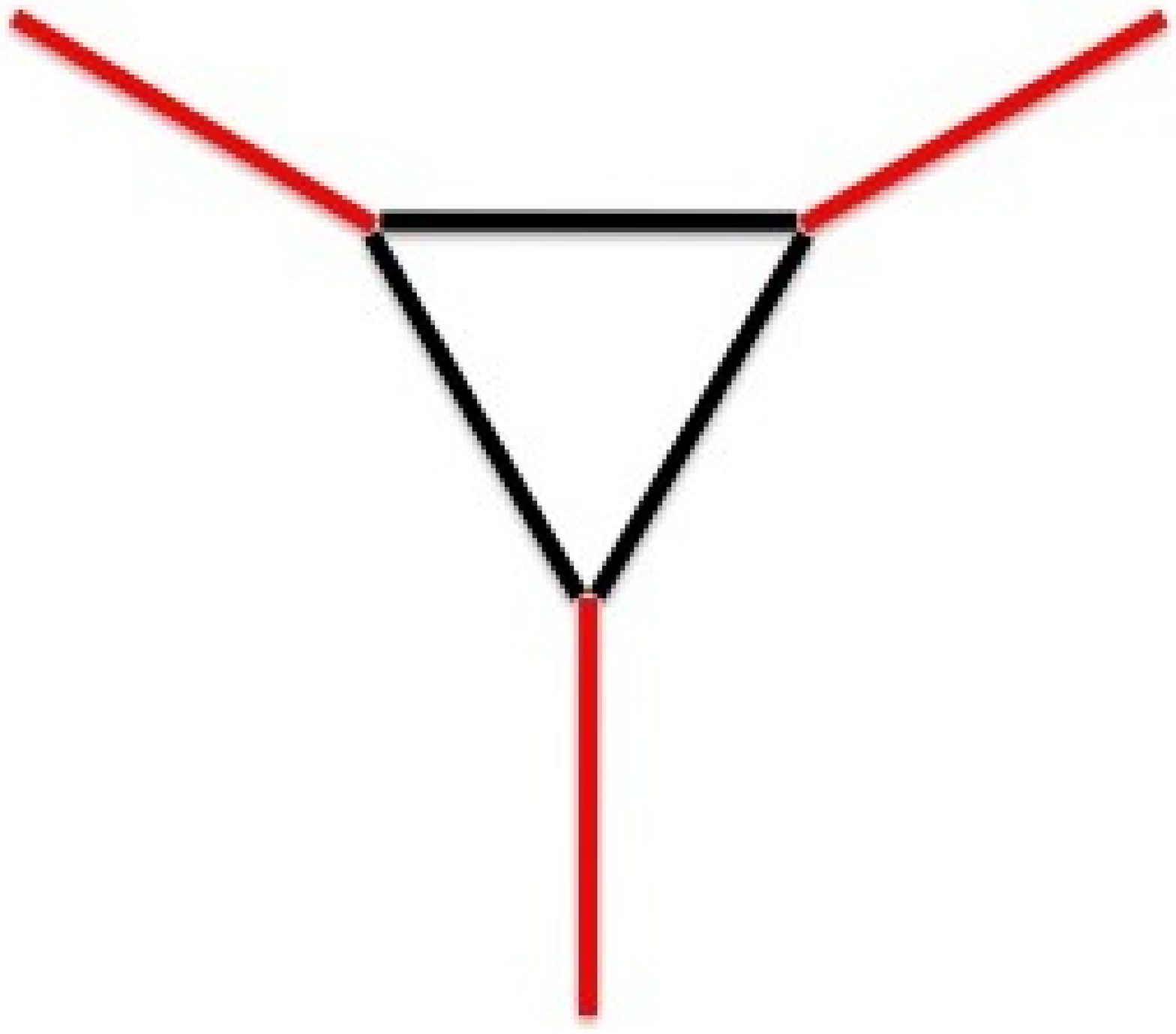}\hspace{2cm}\\
\mbox{}\hspace{2.4cm}(a)\hspace{4cm}(b)\hspace{4.65cm}(c)
\caption{(in colour) (a) On the 3-12 lattice, the three outer edges touched by the same triangle must all have different colours. Given the colours on the outer edges, the inner edges can only be coloured in one way. (b) Mapping to a 3-colouring on the honeycomb lattice, obtained from (a) by removing the inner edges and connecting the outer ones. (c) (See \ref{bounds-3-12}.) A dimer covering on the 3-12 lattice that contains triangles without dimers (as shown here; edges occupied by dimers are red, unoccupied edges are black) can not be made into a 3-colouring, because the fact that the triangle edges form an odd-length loop makes it impossible to satisfy the colouring constraints when they are coloured with the remaining two colours.}
\label{fig:3-12-colourings}
\end{figure}

The 3-12 lattice is shown in Fig. \ref{fig:lattices}(b). It can be regarded as a ``decorated" honeycomb (hc) lattice, the decoration procedure consisting of replacing every vertex on the honeycomb lattice with a triangle that points in opposite directions (up or down) on the two sublattices. Completely analogous to what we did for the 4-8 lattice, we denote the edges of the triangles as ``inner" edges and the edges connecting neighboring triangles as ``outer" edges, and then we consider the possible 3-colourings of the edges of a triangle and the three outer edges touching it. One easily sees that the only allowed 3-colourings are such that the three outer edges all have different colours, in which case the colouring of the inner edges is unique (namely the colour of an inner edge must be the same as that of the outer edge with which it does not share a vertex) [Fig. \ref{fig:3-12-colourings}(a)]. We can therefore forget about the inner edges and so a local 3-colouring on the 3-12 lattice maps to a local 3-colouring on the original honeycomb lattice formed by connecting the outer edges as shown in Fig. \ref{fig:3-12-colourings}(b). Thus, as the number of 3-colourings is the same on the two lattices, while the number of vertices differ by a factor of 3, we get
 \be
W_{\rm{3-12}} = W_{\rm{hc}}^{1/3} = \left(\frac{\sqrt{3}}{2\pi}\right)^{1/3}\sqrt{\Gamma(1/3)}=1.06522\ldots,
\label{exact-3-12}
\ee
where we made use of the exact result for $W_{\rm{hc}}$ \cite{baxter-70}.

\section{Final remarks}
\label{conc}

As noted in Sec. \ref{intro}, a $q$-edge-colouring problem on some lattice can be reformulated as a $q$-vertex-colouring problem on a different but related lattice. In particular, the 3-edge-colouring problem on the 4-8 (3-12) lattice is equivalent to the 3-vertex-colouring problem on the ``square kagome" (``triangular kagome") lattice \cite{kagome-like}. The number of 3-vertex-colourings is in turn equal to the number of ground states in the 3-state AF Potts model on the same lattice. Defining $s$ as the ground state entropy per vertex of the Potts model then gives (sk/tk = square/triangular kagome lattice) 
\be
s_{\rm{sk}} = \frac{2}{3}\log W_{\rm{4-8}}, \quad s_{\rm{tk}} = \frac{2}{3}\log W_{\rm{3-12}}.
\ee 

It is known that some edge-colouring problems are equivalent to \textit{fully packed} loop (FPL) models (or variants thereof) on the same lattice. The 3-edge-colouring problem on the honeycomb lattice can be mapped to an FPL model \cite{baxter-70}, while the 4-edge-colouring problem on the square lattice can be mapped to a fully packed \textit{double} loop (FPL$^2$) model \cite{kh-95}. The 3-edge-colouring problem on the 4-8 lattice can also be mapped to an FPL model [see Eq. (\ref{fp})]; the mapping is identical to that for the honeycomb lattice. In these mappings the loops have fugacity $n=2$. 
In Ref. \cite{jac-99} it was shown that the $n=2$ FPL model on the 4-8 lattice can be mapped to a particular case of a square-lattice $O(2)$ model that is further mappable to the $n=3$ CPL 
model \cite{bl-ni-89}. This route via an FPL and $O(2)$ model therefore constitutes an alternative way of mapping the 3-edge-colouring problem on the 4-8 lattice to the $n=3$ CPL model on the square lattice. The direct mapping given in Sec. \ref{cpl} is however simpler. 

Another example of a colouring problem that can be mapped to a CPL model is the 3-edge-colouring problem on the 4-6-12 lattice, which is another Archimedean lattice with coordination number $q=3$. It can be regarded as a ``decorated" version of the kagome lattice, in which each vertex of the kagome lattice has been replaced by a square, in exactly the same way as when creating the 4-8 lattice from the square lattice (like the square lattice, the kagome lattice also has coordination number 4). Thus ``locally" the 3-colorings are the same as for the 4-8 lattice. Therefore it follows from the analysis in Sec. \ref{local} and \ref{cpl} that there are 18 possible local 3-edge-colouring configurations on the 4-6-12 lattice as well, and the 3-edge-colouring problem on this lattice can be mapped to a CPL model on the kagome lattice with loop fugacity $n=3$. 

\section*{Acknowledgements}

I thank Bernard Nienhuis and Ole Warnaar for an interesting discussion, and an anonymous referee for pointing out Ref. \cite{jac-99}. This work was supported in part by the Australian Research Council. 

\appendix

\section{On bounds for $W_{\cal L}$}

\subsection{The 4-8 lattice}
\label{bounds-4-8}

It is instructive and useful to check the exact result (\ref{W-res}) for $W_{4-8}$ against some lower and upper bounds. We first consider a lower bound that can be derived from the CPL model on the square lattice. Since for positive $n$, 
${\cal Z}(n)$ is a monotonically increasing function of $n$, one can get a lower bound for $W_{\rm{4-8}}$ by replacing $Z_{4-8}={\cal Z}(3)$ with ${\cal Z}(1)$ in the definition (\ref{def-W}). Note that ${\cal Z}(1)$ is just the total number of CPL configurations on the square lattice, which equals $2^M$ since there are two possible local CPL configurations (Fig. \ref{CPL-vertices}) which can be selected independently at each vertex. Since $M=N/4$ this gives a lower bound of $2^{1/4}\approx 1.189$ for 
$W_{\rm{4-8}}$.

An alternative and (as it will turn out) slightly better lower bound for $W_{\rm{4-8}}$ can be established by regarding an arbitrary 3-edge-colouring on the 4-8 lattice as a configuration of fully packed loops on this lattice. To see this, imagine trying to make a 3-edge-colouring by first colouring all edges that have, say, colour A. These edges constitute a dimer covering of
the 4-8 lattice since each vertex must be touched by an A-coloured edge. The remaining edges will form loops on the 4-8 lattice. At each vertex, two edges are covered by a loop while the third edge is ``empty" (i.e. has colour A). As each vertex is visited by one loop, this is a configuration of fully packed loops \cite{nien-lh08}. Since the 4-8 lattice is bipartite, all loops will cover an even number of edges, and so it is possible to colour neighboring edges along each loop in alternating colours as $\cdots$ B-C-B-C-B-C $\cdots$, thus indeed giving a valid 3-colouring. As each loop can be coloured in two different ways (which are related by a shift of the colours by one edge along the loop), the number of 3-colourings is given by 
\be
Z_{\rm{4-8}} = \sum_d 2^{{\cal N}_d},
\label{fp}
\ee
where the sum is over all dimer coverings of the 4-8 lattice and ${\cal N}_d$ is the number of loops formed by the edges that are not covered by a dimer. Thus a lower bound for $Z_{\rm{4-8}}$ is $\mathfrak{Z}_{\rm{4-8}}$, the number of dimer coverings on the 4-8 lattice. It follows that a lower bound for $W_{4-8}$ is given by $\mathfrak{W}_{\rm{4-8}}\equiv \lim_{N\to\infty}\mathfrak{Z}_{4-8}^{1/N}$, which equals \cite{wu-review,sal-nag} $\sqrt{1.457\ldots}\approx 1.207$.

The B-coloured edges also constitute a dimer covering, and so do the C-coloured edges. Thus a 3-edge-colouring can be regarded as consisting of 3 non-overlapping dimer coverings, one of each colour \cite{baxter-70}. When two non-overlapping dimer coverings have been deposited, the third and last one is completely determined. Therefore an upper bound for $Z_{4-8}$ is given by $\mathfrak{Z}_{\rm{4-8}}^2$, the number of ways of depositing two dimer coverings completely independently \cite{baxter-70}. Thus $\mathfrak{W}_{\rm{4-8}}^2 = 1.457\ldots$ is an upper bound for $W_{\rm{4-8}}$. 

A slightly better upper bound can be obtained from (\ref{fp}) by replacing ${\cal N}_d$ with its maximum over the set of all dimer coverings $d$ of A-coloured edges. This maximum is obtained when the loops of B- and C-coloured edges are as short as possible (since the total length of all loops is the same for all
dimer coverings). This occurs when each square is covered by such a loop (i.e. the A-coloured edges are then the ``outer" edges connecting neighbouring squares). Thus ${\cal N}_d=N/4$ for this dimer covering, giving $Z_{4-8}<\mathfrak{Z}_{4-8}\cdot 2^{N/4}$. It follows that $\mathfrak{W}_{4-8}\cdot 2^{1/4} = 1.435\ldots$ is an upper bound for $W_{4-8}$. 

Using the best lower and upper bounds established above, we have therefore found that $1.207\ldots < W_{\rm{4-8}} < 1.435\ldots$, which is consistent with the exact result (\ref{W-res}). We see that the exact result is quite close to the lower bound; the ratio between these numbers is $\approx 1.027$. 
This ratio is very similar to the corresponding one for the honeycomb lattice, which like the 4-8 lattice is bipartite with coordination number 3:
Using the results in Refs. \cite{baxter-70} and \cite{wu-review}, the ratio for the honeycomb lattice is $\sqrt{1.460\ldots/1.381\ldots}\approx 1.028$.

\subsection{The 3-12 lattice}
\label{bounds-3-12}

Defining $\mathfrak{W}_{\rm{3-12}}=\lim_{N\to\infty} \mathfrak{Z}_{\rm{3-12}}^{1/N}$ where $\mathfrak{Z}_{\rm{3-12}}$ is the number of dimer coverings on the 3-12 lattice, we have \cite{wu-review,jof} $\mathfrak{W}_{\rm{3-12}}=2^{1/6}\approx 1.122$ which is greater than the exact result for $W_{\rm{3-12}}$ in (\ref{exact-3-12}). Thus for the 3-12 lattice, the number of dimer coverings is not a lower bound for the number of 3-colourings, unlike the situation for the 4-8 lattice. This is because on the 3-12 lattice, there are dimer coverings that cannot be used as a starting point for generating 3-colourings (in the sense discussed in \ref{bounds-4-8}). This is related to the fact that the 3-12 lattice is not bipartite, which leads to the existence of odd-length loops, as illustrated in Fig. \ref{fig:3-12-colourings}(c).


\begin{thebibliography}{99}
\bibitem{elab} For the square lattice the calculation was done for a 24-vertex model that is equivalent \cite{dc-nien-04} to a formulation of the 4-edge-colouring problem in terms of a fully packed double-loop model \cite{kh-95}. Later an algebraic Bethe Ansatz was developed for the square-lattice problem \cite{jac-zj}.
\bibitem{alt-name} Especially the 3-12 lattice also goes by many other names in the literature, including the star lattice, Fisher lattice, expanded kagome lattice, and triangular honeycomb lattice. The 4-8 lattice is sometimes instead called the
square-octagon lattice.
\end{thebibliography}
\end{document}